\documentclass[10pt,conference]{IEEEtran}
\usepackage{cite}
\usepackage{amsmath,amssymb,amsfonts}
\usepackage{algorithmic}
\usepackage{graphicx}
\usepackage{textcomp}
\usepackage{xcolor}
\usepackage{booktabs} 
\usepackage{url}
\usepackage{hyperref}

\newcommand{\rqScorer}{RQ0: How accurate is the Scorer?}
\newcommand{\rqOasisvsServices}{RQ1: How correct is OASIS across services?}
\newcommand{\rqOasisvsBaselines}{RQ2: How does OASIS compare with the baselines?}

\newcommand{\rqOasisvsModels}{RQ3: Are OASIS results model dependent?}
\begin{document}

\title{OASIS: Static and Dynamic OpenAPI Specification Inference}

\author{\IEEEauthorblockN{Hamza Bin Mazhar, Yuqing Wang, Jesse Nyyssölä, Mika Mäntylä}
\IEEEauthorblockA{\textit{Department of Computer Science, University of Helsinki}, Helsinki, Finland \\
\{hamza.mazhar, yuqing.wang, jesse.nyyssola, mika.mantyla\}@helsinki.fi}
}
\maketitle

\begin{abstract}
OpenAPI specifications are crucial in black-box testing. Yet, many services have no specifications, or the specifications are incorrect. Existing tools for specification generation read the source code or the documentation, but do not utilize the dynamic view of the system.
We present OASIS, an OpenAPI specification generation tool that drafts a specification from the static source code and then validates and repairs it dynamically against the running service with an LLM agent. To evaluate specifications without a ground truth, we build a Scorer. It reads a specification as a set of claims and confirms or contradicts each claim with requests to the running service. We evaluate our approach on 30 services in seven frameworks and five languages, and against five baselines from prior work.
OASIS finds 99\% of the routes in the 30 services, and 97\% of the claims that a request can test are true. OASIS produces better quality specifications than the baselines in 52 of 54 cases. It takes about one cent of API cost per route. A human evaluation finds the Scorer correct on 94\% of its outcomes. Where the Scorer and the ground truth specifications disagree, the human finds the Scorer right in 107 of the 117 cases.
Overall, our work shows that a specification generated from the source code and then validated and repaired dynamically with an LLM agent is more complete and more correct than one generated from the source code alone.
\end{abstract}


\section{Introduction}
\label{sec:intro}
OpenAPI specifications (OAS) are used for black-box testing~\cite{arcuri2021evomaster,atlidakis2019restler,hatfielddodds2022schemathesis}, API documentation~\cite{swaggerui2026} and client code generation~\cite{lercher2024evolution,koren2018exploitation}. These tasks require a specification that accurately represents the running service. However, many REST services have no specification~\cite{neumann2021,lazar2025oasbuilder}, and those that exist can drift from the code as the API changes~\cite{lercher2024evolution,decrop2025restspecit,redocly2026drift, singh2026speculate}. 
Some of them have faults due to drift or mistakes which can lower the effectiveness of black-box testing tools~\cite{mazhar2026fault, kim2022rest} or make them crash~\cite{sahin2025wfd}.

It would be ideal if one would get specifications automatically. This could not only reduce errors in specifications, but also ensure that specifications exist in the first place and do not become incorrect due to drift, as maintaining them manually after each change is tedious. Recent work has moved towards this direction. Most specification automation tools read the source code, with program analysis~\cite{huang2024respector,lercher2024autooas,serbout2023expresso} or with an LLM~\cite{singh2026speculate,deng2026lrasgen,chen2026oops,hu2026specweaver}. Others convert existing documentation into an OAS~\cite{lazar2024specrawler,lazar2025oasbuilder,kesraoui2025design}. These tools work from static artifacts and do not perform dynamic validation, i.e., send requests to the running service, so they cannot check their claims against the replies of the service. They may miss behavior that only appears at runtime. For example, a handler may return a database model without naming any of its fields or the framework adds replies that no handler writes.%

Existing tools evaluate their specifications against a ground truth, which is built by hand~\cite{huang2024respector,singh2026speculate,deng2026lrasgen,chen2026oops} or shipped with the service~\cite{serbout2023expresso,yandrapally2023carving}. They count every difference from the ground truth as an error of the tool, even though a ground truth can be wrong or outdated~\cite{huang2024respector,deng2026lrasgen}.
Many services also have no ground truth.

To close gap 1, \textbf{the lack of dynamic validation of specifications}, we present OASIS (OpenAPI Specification Inference from running Services). To our knowledge, OASIS is the first tool that infers an OAS from source and then validates and repairs it against a running service using an LLM agent.

To close gap 2, i.e. \textbf{the dependence on a reliable ground truth to compare the generated specification against}, we build a Scorer that reads a specification as a set of claims. It builds the requests from the specification to evaluate and then uses the replies to see if they confirm or contradict each claim. In order to evaluate the validity of this Scorer, we first evaluate the Scorer by hand and find it correct on about 94\% of its outcomes. Where the Scorer outcome and the ground truth evaluation outcome disagree, the ground truth is wrong in 107 of the 117 cases.

We evaluate OASIS on 30 services in seven frameworks and five languages, against five baselines: Respector~\cite{huang2024respector}, Speculate~\cite{singh2026speculate}, ExpressO~\cite{serbout2023expresso}, LRASGen~\cite{deng2026lrasgen} and OOPS~\cite{chen2026oops}. We use GPT-5.6 Luna for OASIS and the LLM-based baselines. We also run OASIS with seven low-cost models from six providers, in order to see how its results depend on the model.
OASIS leads all five baselines on the number of routes it can find, and on the status codes and fields it documents. It costs about one cent of API cost per route. We evaluate the following research questions: 
\begin{itemize}
\item \rqScorer
\item \rqOasisvsServices
\item \rqOasisvsBaselines
\item \rqOasisvsModels
\end{itemize}

\section{Background}
\label{sec:background}
\subsection{OpenAPI Specifications}
\label{sec:background:oas}

An OAS is a JSON or YAML document that lists the paths of a REST API and the operations on each path~\cite{openapi2024specification}. A route is an HTTP method and a path that a service serves, such as \texttt{PUT /api/playlist-folders/\{playlistFolder\}} in Fig.~\ref{fig:openapi-example}. A specification documents each route as an operation. For each operation it states the parameters, the request body and its schema, the response status codes with their bodies, and the security scheme. The function that the framework calls for a route is its handler. The list of routes that a service registers is its route registry. 

\subsection{Claims of a Specification}
\label{sec:background:claims}
\begin{figure}[t]
\centering
\includegraphics[width=\columnwidth]{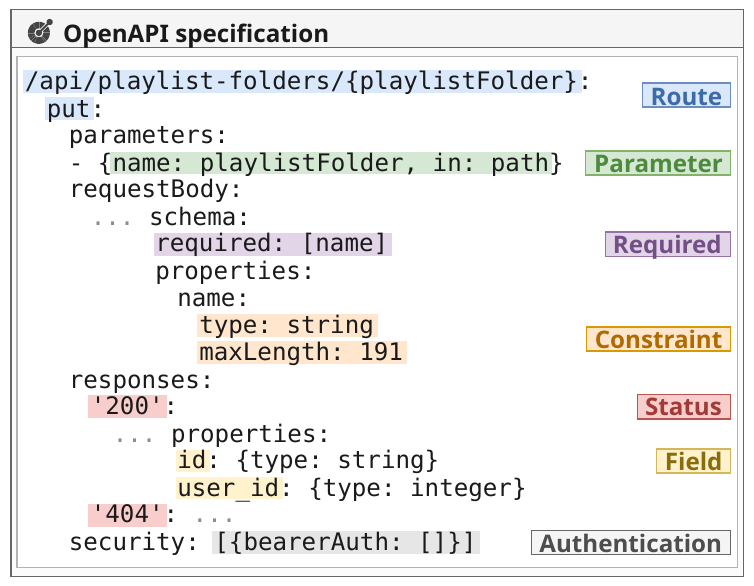}
\caption{An operation from the Koel service.}
\label{fig:openapi-example}
\end{figure}

An OAS can be read as a set of claims. A claim is one statement that a request to the running service can confirm or contradict. For example, Fig.~\ref{fig:openapi-example} shows an operation from a service called Koel, a music streaming server. The operation claims that the route exists and takes the folder as a path parameter. It claims that the body must contain \texttt{name}, a string of at most 191 characters. It claims that the operation answers 200 and 404, that a 200 reply carries \texttt{id} and \texttt{user\_id}, and that a request needs a bearer token. In this order, the seven kinds of claims of an OAS are \textbf{Route}, \textbf{Parameter}, \textbf{Required}, \textbf{Constraint}, \textbf{Status}, \textbf{Field} and \textbf{Authentication}.

A testing tool that reads the specification works from these claims and trusts them. If the maximum length is wrong, its boundary tests are wrong too. If it claims a route that does not exist, the testing tool sees a 404. The quality of the specification matters for black-box testing because in most black-box testing tools, their sole input is an OAS~\cite{golmohammadi2023survey,arcuri2021evomaster,atlidakis2019restler,hatfielddodds2022schemathesis}. The quality also matters for client code generation and documentation, where the specification needs to match the service and its latest changes~\cite{lercher2024evolution,koren2018exploitation}. A developer who reads the generated documentation trusts the claims. If the 404 reply is missing, the developer's code does not handle it. A client generated from the specification is built from these claims too. If \texttt{user\_id} is missing from the 200 reply, the generated client has no field for it. 

\section{Related Work}
\label{sec:related}

Recent studies have focused on automated generation of OpenAPI specifications from source and documentation, and also from UI and traffic. Most of these works evaluate their specifications against some ground truth specification. Testing-related tools use generated specifications to generate test cases or assess responses.

\subsection{Specification Generation From Static Sources}
\label{sec:related:static}

Many tools write OpenAPI specifications from the source without sending the service any request. Respector~\cite{huang2024respector}, AutoOAS~\cite{lercher2024autooas}, ExpressO~\cite{serbout2023expresso}, Speculate~\cite{singh2026speculate}, and SpecWeaver~\cite{hu2026specweaver} use static analysis to parse the handlers. The authors have noted that static analysis can miss the routes that are handled dynamically \cite{huang2024respector} and deep levels of nested schemas are handled as plain objects~\cite{singh2026speculate}. The findings of the static analysis can be fed to an LLM to create a specification such as in LRASGen~\cite{deng2026lrasgen}. As in OOPS~\cite{chen2026oops}, LLMs can also be used to create an agent workflow for extracting a dependency graph that can further be used to create and refine the specification. Some tools such as OASBuilder~\cite{lazar2025oasbuilder}, SpeCrawler~\cite{lazar2024specrawler} and AutoREST~\cite{cao2017autorest} can scrape HTML documentation to create the specification. This approach can have low recall on deeply nested responses and in cases of poor existing documentation~\cite{lazar2025oasbuilder}. Furthermore, the specifications built from documentation may carry forward the errors of existing documentation. None of these tools send any request to the running service as part of their pipeline for specification generation. The LLM-based tools are susceptible to hallucinations, i.e., they can invent routes or parameters, and without a request to the running service, these claims cannot be validated.

\subsection{Specification Generation From Traffic and User Interfaces}
\label{sec:related:traffic}

Another category of tools uses the traffic of a running service to generate API specifications. Tools like SpyREST~\cite{sohan2015spyrest} and Map Reduce Graph (MRG)~\cite{levi2026mrg} use HTTP traffic to create REST API documentation, and APISensor~\cite{yang2026apisensor} uses it to discover the endpoints of an API. Some approaches use call examples to infer specifications~\cite{eddouibi2017example}. APICARV~\cite{yandrapally2023carving} runs UI tests of a web application and records the calls to build a graph of the paths it saw. It then probes the service to fill any gaps in the graph to create an OAS. FOREST~\cite{jeong2026forest} uses manual user interactions, with mutated requests to create an OAS. Practitioners also use tools like mitmproxy2swagger and Optic~\cite{mitmproxy2swagger2026, optic2026} to create specifications from captured traffic. The specifications produced by these tools are only as complete as the traffic behind them. If the traffic misses an interaction, it does not enter the specification.

RESTSpecIT~\cite{decrop2025restspecit} starts from a name of a public API and then asks an LLM to guess requests from its own training knowledge. It sends these requests and keeps the accepted claims. Users can also add request seeds manually, but that requires effort. Without seeds, its route recall can be low.

REST testing tools like EvoMaster~\cite{arcuri2019evomaster}, RESTler~\cite{atlidakis2019restler}, Schemathesis~\cite{hatfielddodds2022schemathesis} and Morest~\cite{liu2022morest} do send requests to the service, but their main goal is to test the service. They start with an existing OAS and during testing, most of them do not modify the specifications with what they learn. However, some testing tools like NLP2REST~\cite{kim2023nlp2rest} and RESTGPT~\cite{kim2024restgpt} do append the specification with rules, examples and values that they gather from the natural-language descriptions of the specifications. NLP2REST validates these with actual requests to the service. Both of these tools need a specification with the description fields to start with. SAINT~\cite{pan2026saint} requires no OpenAPI specification and produces route models for Java using static analysis and an LLM. It uses an LLM agent to repair the requests that draw a 4xx from the running service, but its output is Java tests, not an OpenAPI specification.

\subsection{Specification Evaluation}
\label{sec:related:evaluation}

Most specification generation tools evaluate the quality of the specification against a ground truth. This ground truth is sometimes hand built or labeled, which is the case for Respector~\cite{huang2024respector}, LRASGen~\cite{deng2026lrasgen}, OOPS~\cite{chen2026oops}, AutoOAS~\cite{lercher2024autooas}, Speculate~\cite{singh2026speculate} and OASBuilder~\cite{lazar2025oasbuilder}. ExpressO~\cite{serbout2023expresso} and APICARV~\cite{yandrapally2023carving} use the developer-published specifications that already come with the services. A hand written ground truth itself can miss information that is only available at runtime or be wrong. Respector found 228 routes and 502 responses missing from the developer-written specifications of its subjects~\cite{huang2024respector}. The ground truths of Respector were later fixed by AutoOAS in its study~\cite{lercher2024autooas}, after its authors found wrong paths and missing prefixes. RESTSpecIT~\cite{decrop2025restspecit} found working routes missing from official documentation. Therefore, a tool that produces a specification that matches the service more closely than a ground truth specification would be negatively evaluated due to its difference from the ground truth. Some tools like Schemathesis~\cite{hatfielddodds2022schemathesis}, Dredd~\cite{dredd2026} and Prism~\cite{prism2026} validate the replies they get against the specification, but they leave the specification unchanged.

\subsection{Positioning OASIS}
\label{sec:related:motivation}

The specification tools above see only certain parts of a service. The tools that read the source cannot see what the service shows at runtime, and the tools that learn from the traffic and user interaction can only see what the traffic can produce. OASIS uses both: the static source to create a draft specification, and the runtime replies and behavior to dynamically validate it. To evaluate the OASIS specifications, we build a Scorer that generates requests from a specification and gives an outcome to every claim that a request can decide. We can then compare the specifications from different tools and OASIS without the need for a ground truth specification. This also helps evaluate any specification if a ground truth is absent for a service.

\section{OASIS Methodology}
\label{sec:methodology}

\begin{figure*}[t]
\centering
\includegraphics[width=\textwidth]{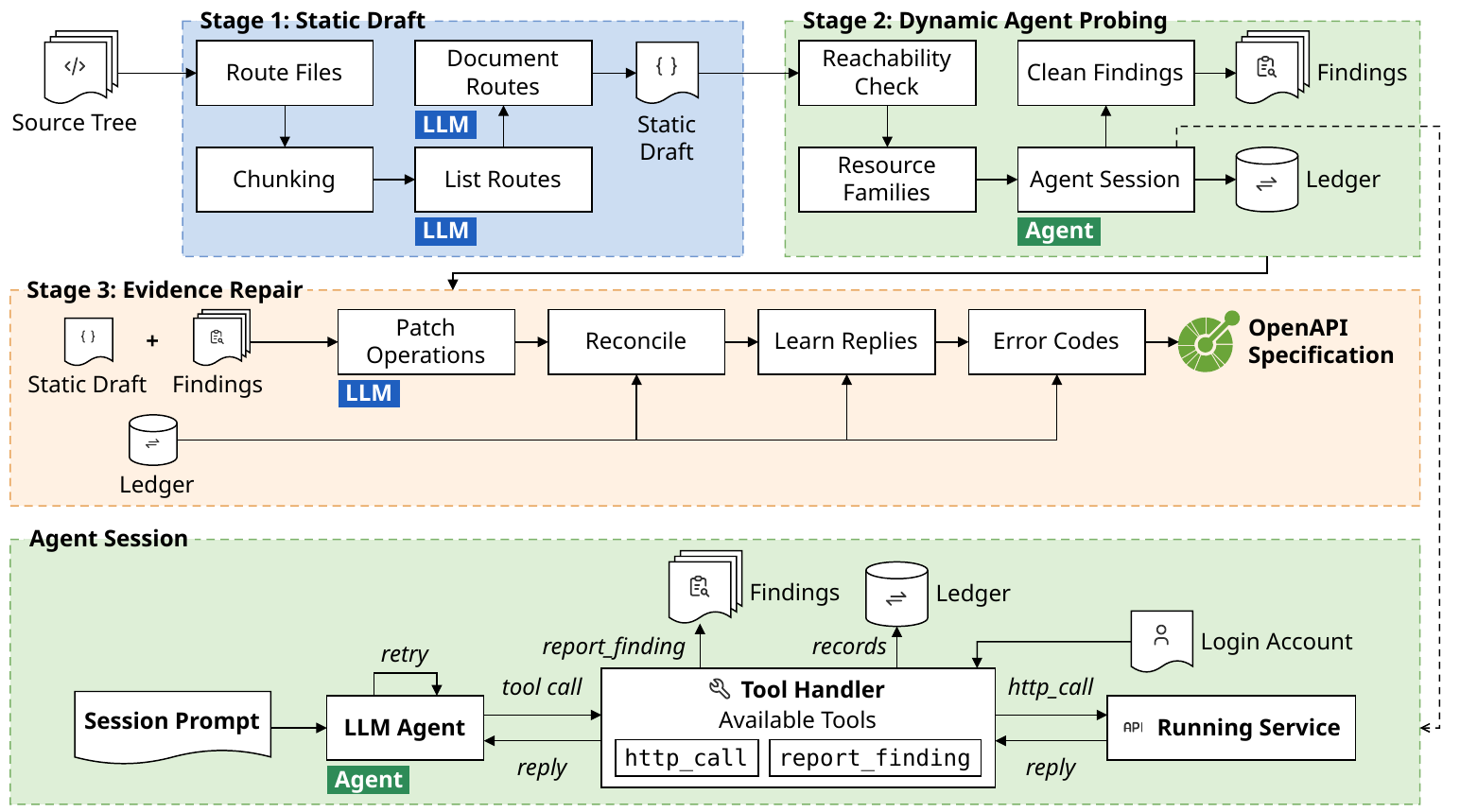}
\caption{OASIS pipeline, the lower panel opens the Agent Session of stage 2.}
\label{fig:pipeline}
\end{figure*}

Fig.~\ref{fig:pipeline} shows the OASIS pipeline. The three inputs for OASIS are the Source Tree of the code, the URL of the Running Service and a Login Account. OASIS creates an OpenAPI specification in three stages. In stage 1, OASIS reads the static source code and creates a Static Draft specification. In stage 2, an LLM Agent probes the service with the requests generated from the Static Draft. OASIS stores the Findings, which record where the replies from the service disagree with the Static Draft, and a Ledger of every request and reply for the next stage. In stage 3, OASIS uses the Findings and the Ledger to validate and repair the Static Draft. The result is the final OpenAPI Specification.

\subsection{Stage 1: Static Draft}
\label{sec:methodology:static}

Stage 1 takes the Source Tree as input and outputs the Static Draft just from static source alone. 

\textbf{Route Files:} OASIS first finds the files that register routes. It parses each source file into a syntax tree using tree-sitter~\cite{brunsfeld2026treesitter}. It skips test files, build outputs, and third-party code that is not directly related to the service implementation, like node modules and vendor folders. A selector script then goes through the annotation and call nodes of the tree, looking for framework registration markers, such as \texttt{@GetMapping} in Spring or \texttt{router.get} in Express. Since the selector only goes through these nodes, any marker in a string or a comment is skipped. We use tree-sitter because it has grammars available for all the languages of the benchmark, and it does not need a build of the service. A text or regex script here would return the markers in comments and strings, which can increase the LLM calls and API cost. Every file that has a marker becomes a Route File.

\textbf {Chunking:} The LLM requires request types and validation rules of a route to document it, but the whole Source Tree can be too large to fit in just one prompt. OASIS builds a chunk from each Route File, the files it imports, and the files those import in turn.
Files are put in the chunks in the order of need. The Route Files are first and are never cut, followed by files that define request types or validation rules. Then the other imported files and helpers are added. OASIS cuts the type and validation files at 24,000 characters and the imported files at 12,000 characters. Any file that does not fit the budget of 60,000 characters is removed, with the least useful files dropped first. 

\textbf{List Routes:} The LLM takes these chunks as inputs to list their routes. Since the LLM may miss some routes from the files, OASIS also counts routes using pattern-matches. The pattern-matched list is built using a regular expression per framework, and its count acts as a lower bound for the LLM to build the route list against. If the LLM's listed route count falls short of the pattern-matched count, the LLM is told to try again for up to five attempts. If it still falls short of the lower bound, the pattern-matched route locations in code are shown to it. The pattern-matched list is not given in the beginning since it may mislead the LLM down paths which are not necessarily a route. Furthermore, simple pattern-matching cannot add the class or group prefix that the LLM is able to add. 

\textbf{Document Routes:}
The LLM is then prompted to document each route with the parameters, request body, constraints, and responses. In some cases, the source can show the responses directly in the code, and the next stages can fill the rest. The LLM gets another try if it does not return any responses, or body fields for a POST, PUT or PATCH, or any parameter for any other method. The two answers get merged and one last call to the LLM asks for anything that the two missed. This forms the Static Draft.

\subsection{Stage 2: Dynamic Agent Probing}
\label{sec:methodology:probing}
Stage 2 takes the Static Draft of stage 1 and a Login Account, and then runs an LLM Agent on each Resource Family to dynamically validate it against the Running Service. It produces two things: Findings and a Ledger.

\textbf{Reachability Check:} It first checks whether all the listed routes in the Static Draft are reachable, and if any are not, they are removed. The check only sends GET and OPTIONS requests, and changes no status code, field or parameter.

\textbf{Resource Families:} OASIS groups routes sharing the same first path segment into Resource Families. Nested paths remain under a parent since they might need the parent's ID. Read-only families run before write families to avoid interference.

\textbf{Agent Session:} One Agent Session runs for each Resource Family, with four running in parallel. 
The Session Prompt enlists: the routes in a family in lifecycle order (creates first, deletes last), the methods each path allows from its \texttt{Allow} header, the IDs that may have been known from earlier replies, the draft operations and the handler source. We give the agent two tools. The first one is \texttt{http\_call} which sends a request that the agent builds from the Static Draft. The second one is \texttt{report\_finding} which records disagreement between the Static Draft and the service. Both of these tools run through the Tool Handler. We include some strict safety rules for the Tool Handler. It checks each request to see that it follows the rules. The rules entail that a DELETE will only go through for a resource created within the session and no request is allowed to change the login account credentials. Since some requests may change service state, we make sure that the copy of the service that OASIS sees is able to be discarded without any consequences. When a request logs the account out, later requests can then lead to 401 or 403 status codes. The Tool Handler then logs back in and repeats the refused request once.

The Tool Handler hands the replies back to the agent, which compares them with the Static Draft and reports the differences with \texttt{report\_finding}. The Tool Handler stores these Findings with the matching request/reply sequence as evidence. Every request and reply also goes into the Ledger.

The agent divides these Findings into seven categories. Three categories relate to the requests: the service ignores a rule, contradicts a required mark or enforces a rule missing from the specification. Three categories relate to the replies: a missing or wrong status code, a response body or error bodies that have a different shape in the specification. One relates to a parameter where a service shows a different type, format or location of a parameter than the one in the specification. The agent first tries to get a successful response from the service, using requests a service would accept. After that, it takes the accepted requests and starts testing them one claim at a time, mutating the claims to see if the service still accepts the requests. For example, if a required field is left out in a request, and the service accepts it, the required field was not actually required. The accepted requests from this stage become part of the examples of the specifications. A session gets at most 25 HTTP calls, 60 model calls and 900 seconds. 

\textbf{Clean Findings:} After the sessions end, the Findings are cleaned. Contradicted findings are dropped and duplicate findings are merged. OASIS also merges the same 5xx server errors on three or more routes into a single finding, since they point to one fault in the service. At the end, OASIS deletes the resources that were created during the Agent Sessions, and the Ledger contains every request with its reply for the next stage. The Findings and the Ledger are the two outputs of stage 2.

\subsection{Stage 3: Evidence Repair}
\label{sec:methodology:repair}
Stage 3 takes the Static Draft, the Findings and Ledger from stage 2 and delivers the final OpenAPI Specification. It uses the evidence that stage 2 collected and uses that to edit the draft. An addition of a claim needs a reply that shows it, and a removal of a claim requires a contradiction. 

\textbf{Patch Operations:} To do these edits, the LLM writes one JSON patch for each operation that has any finding. It writes the patch using the available information in the operation, its findings, and the handler source. The \texttt{report\_finding} tool asks the agent to name a part of the operation for which there was a finding, such as a field or a status code.
Using this, OASIS can ensure that the patch is applied only to the named part. Any patch that leaves the operation malformed is rejected. A patch can add claims, but the rules ensure that it cannot remove a claim that no finding contradicts.

\textbf{Reconcile:} Afterwards, OASIS reconciles the specification with the Findings and the Ledger. Any success code that is contradicted is changed to the one that the service actually returned. OASIS marks a field required when sending a request without the field drew a 4xx. If the request still draws 2xx even when the required field is not sent, then the field is unmarked as required. OASIS does not let server errors change anything in the specification.

\textbf{Learn Replies:} Then OASIS learns from the replies in the Ledger of stage 2. This step makes no LLM calls. For each operation and success code, OASIS reads up to 20 bodies that the service returned with the code. We then add response fields that these bodies carried, and drop the required mark of a field that one of the bodies left out. We write request examples from the accepted requests, leaving out the IDs and the account's own data. We also declare fields that the service accepted, by checking if their values or names appear in replies or validation errors. We also add an operation's \texttt{security} entry according to whether the request from the agent required a credential or not to be accepted by the service.

\textbf{Error Codes:} Lastly, OASIS also adds the 4xx error codes the service returned to each operation. It leaves out the codes that the agent might have caused itself such as 405 (method not allowed) and 429 (too many requests). OASIS also removes the operations under a leading path segment which always drew a bare 404 or a 405. A bare 404 or 405 is the router actually replying to a path that no handler serves. The final OpenAPI Specification is the output of stage 3.

\section{Evaluation Method}
\label{sec:setup}

We evaluate OASIS using four research questions. The results of the evaluation depend on the accuracy of the Scorer, so we first evaluate that in RQ0. RQ1 then measures OASIS using the Scorer. RQ2 compares OASIS to the baselines. The baselines do not send any request to the service, so RQ2 also shows what dynamic validation with probing and repair adds on top of reading the source. RQ1 and RQ2 use one LLM, and RQ3 evaluates how much the results depend on the LLMs.

\subsection{Benchmark and Deployment}
\label{sec:setup:benchmark}

Table~\ref{tab:services} shows 30 open-source services we use in this evaluation. The benchmark is spread over seven frameworks and five languages. In total there are about 700,000 lines of code (LOC) and 2213 registered routes. 16 services in this benchmark are from earlier studies on specification generation~\cite{chen2026oops, deng2026lrasgen} and the EvoMaster Benchmark~\cite{arcuri2023emb, sahin2025wfd}. We add four RealWorld backends~\cite{realworld2026demo} that implement one shared API in different frameworks, and 10 further applications. We found most of the added services by searching GitHub per framework, and we preferred applications from different domains over templates and tutorials. We included a service in the benchmark when we could build and run it locally, when it needed no login or accepted one account, and when it had a route registry. We read the registry from the framework's route table, start-up output or the running router. If none is available, we analyze the route registrations in the source. Nine services need no login, while the other 21 have one account which is either user, admin, or owner, whichever a service grants. Each group in Table~\ref{tab:services} has five services. 20 services have a ground truth (GT), out of which 18 are OpenAPI documents.

\subsection{Baselines}
\label{sec:setup:tools}

We compare OASIS with five baselines, Respector~\cite{huang2024respector}, Speculate~\cite{singh2026speculate}, ExpressO~\cite{serbout2023expresso}, LRASGen~\cite{deng2026lrasgen} and OOPS~\cite{chen2026oops}. Speculate, LRASGen and OOPS are three recent tools that read the source using an LLM. Respector and ExpressO use no LLM. All five baselines take the Source Tree as input, just like stage 1 of OASIS. Each baseline runs on every service that the tools support. The only exception is LRASGen due to its high cost. It runs on the two largest services of each framework it supports determined by registered routes count. We do not include RESTSpecIT~\cite{decrop2025restspecit} or the traffic-based tools, since neither fits our benchmark. RESTSpecIT creates its requests from what an LLM knows about the API through its training, and it found only 59\% of the routes of APIs that the LLM did not know. The traffic-based tools need recorded traffic.

\subsection{Runs, Models and Cost}
\label{sec:setup:runs}

We use OpenRouter as the LLM provider for OASIS and all the baselines that require an LLM. The LLM is set to zero temperature for OASIS to keep the output as repeatable as possible by API calling. We do not set the reasoning parameter explicitly, which means that the model reasons at its provider's default. The baselines use their own prompts and temperature, which they mention in their papers. We run them from their public repositories and change nothing other than what is needed to build, start and reach the model API. A reset script exists to restore the accounts before every scoring run. We run OASIS and the baselines on GPT-5.6 Luna, which RQ3 finds to be the best model. We run OASIS three times over the entire benchmark for RQ1 to see if results are repeatable. For RQ2, the baselines are run once on GPT-5.6 Luna. A baseline run that fails is retried once, and if it fails again we stop it. We compare the three runs of OASIS separately with the baseline and then take the mean. For the LLM evaluation in RQ3, we run OASIS using seven LLMs from six providers on the largest service by registered routes count in each framework. Each model is from its provider's low-cost tier at the time of writing, supports function calling and has a context window of at least 128K tokens. The set covers GPT, Llama, Gemini, DeepSeek and Qwen, and GLM for breadth.

Cost is calculated from the price in USD that the provider API reports per call. The list price of GPT-5.6 Luna was 0.20 USD/million input tokens and 1.20 USD/million output tokens at the time of writing this paper.

\subsection{Scoring Against the Running Service}
\label{sec:scoring}

We use our Scorer to evaluate the specification against the running service. It checks for the seven kinds of claims mentioned in Section~\ref{sec:background}. It exclusively uses the specification to create requests, only taking IDs from the service's own data, and then validates whether the specification is correct or not about these claims. A reply from a service that agrees with the claim confirms it, and a disagreement contradicts it. A claim that cannot be confirmed or contradicted remains undecided. 

The Scorer logs in with the service's account and uses those credentials in every request it sends, except the requests that test an authentication claim. A \textbf{route} claim is confirmed when the reply to the Scorer's request shows that the route exists. A \textbf{parameter} claim is confirmed when the Scorer's request that carries the parameter is accepted with a 2xx. The Scorer never contradicts a parameter claim because a reply is not enough to decide whether the service used the parameter or ignored it. For a \textbf{required} and \textbf{constraint} claim, the Scorer changes one thing in an accepted request, such as leaving out a required field, or setting a value outside a constraint. If the unchanged request was first accepted, but now returns 4xx, this means that these claims are confirmed, and a 2xx contradicts them. A \textbf{status} claim is confirmed when the service's reply returns that status code. For success codes, the Scorer also confirms if the response body matches the schema in the specification. A \textbf{field} claim is confirmed when a successful reply from the service carries the field along with its listed type. An \textbf{authentication} claim is confirmed when the service refuses a request without the valid credentials, and is contradicted when the service allows a request without the credentials. 

Since we use the Scorer for evaluating all specifications in later results, we first evaluate its accuracy in Section~\ref{sec:human-evaluation}. To do this, we compare the outcomes it produces for the claims to what the ground truth specifications say about the claim. These outcomes can either agree, disagree or the ground truth specification says nothing about the claim. In the case of disagreement, a human manually evaluates the claims. This becomes the human's outcome on these claims. For example, if a specification marks a field as required, and the Scorer confirms it, but the ground truth specification says that this field is not required, the human evaluates whether the field is required or not. The human also evaluates 100 claims on which the Scorer outcome and the ground truth specification agree for control, to see if both are correct or wrong. Since this evaluation requires a ground truth specification that is a valid OpenAPI document, we do this for 18 services in our benchmark that have one. The human evaluates every confirmed and contradicted claim that the ground truth specifications say nothing about. The only exceptions are confirmed status codes and fields. These are too many to evaluate, so we sample them. In total, the human evaluates 1865 outcomes. The evaluation is blind, so the human does not know what the ground truth specification or the Scorer say about a claim. The tools that produced the specification claims are hidden as well. The human analyzed the source code implementation and used curl requests to decide the claims manually. A second human also re-evaluated a random 5\% of these 1865 outcomes to see if they repeat.

For RQ0, we report the agreement of the human outcome, Scorer outcome and ground truth specification to show the Scorer error.

\subsection{Evaluation Metrics}
\label{sec:metrics}

We report the following metrics for all specifications from RQ1 to RQ3 using the Scorer. In every RQ, we compare the specifications for one service: Static Draft and the final OpenAPI Specification in RQ1, OASIS and the baseline specifications in RQ2, and the specifications of seven models in RQ3.

\begin{itemize}
\item \textbf{Precision} counts how many claims the service confirms out of the claims that a request can decide.
\item \textbf{Tested share} counts the claims that can be confirmed or contradicted with a request from the Scorer out of all the claims of a specification. We leave out writes that cannot be decided, such as file uploads. The other claims need other roles or data that the service does not hold, or no request from one account reaches them. The precision above is over this share.
\item \textbf{Invented routes} counts the routes in the specification which are not in the route registry, so fewer is better.
\item \textbf{Route recall} counts the number of routes present in the specification out of the total routes in the route registry.
\item \textbf{Status recall} counts the status codes that the specification has for an operation out of all the status codes returned by the service for that operation. We do not count status codes such as 403 and 429, redirects and server errors because they depend on the account, the client or a fault.
\item \textbf{Field recall} counts the fields of the service's reply that the specification has for an operation and a success code. We use the top-level fields for this measure. 
\end{itemize}

For both status and field recall, we use the replies to the requests of all the specifications in the comparison. This means that the status codes and fields of a specification are evaluated against a wider reference than its own replies.

\begin{itemize}
\item \textbf{Claim F1.} The status and field recall are enough for RQ1 and RQ3, where every specification comes from OASIS and states every kind of claim that its service can have. However, when comparing against specifications of different tools like in RQ2, and especially for claims that the replies cannot show, we use claim F1. This is a measure that takes all the confirmed claims of all the specifications being compared as its reference. For each of the seven kinds of claims, we take the harmonic mean of precision and relative recall, and claim F1 is their mean. Relative recall for a kind of claim counts the claims the specification states, out of the claims the service confirmed for any of the compared specifications.
\item \textbf{GT agreement} compares a specification with the ground truth specification of its service, as earlier tools do~\cite{huang2024respector,singh2026speculate,deng2026lrasgen,chen2026oops}. It is the mean F1 over operations, parameters, constraints and responses. We treat it as a secondary measure.
\end{itemize}

\section{Results}
\label{sec:results}

\subsection{\rqScorer}
\label{sec:human-evaluation}

Table~\ref{tab:human-evaluation} shows the result of human evaluation. An outcome is the Scorer's verdict on one claim which can be either confirmed or contradicted.

\begin{table}[t]
\centering
\caption{The human evaluation of the Scorer. Rows marked * are sampled. Error is Scorer wrong over Human Evaluated.}
\label{tab:human-evaluation}
\footnotesize
\setlength{\tabcolsep}{3pt}
\begin{tabular}{l rrrr}
\toprule
 & Outcomes & \shortstack{Human\\Evaluated} & \shortstack{Scorer\\wrong} & \shortstack{Error\\(\%)} \\
\midrule
\multicolumn{5}{l}{\textit{Scorer vs GT}} \\
\hspace{1em}Agree$^*$ & 6899 & 100 & 3 & 3.0 \\
\hspace{1em}Disagree & 117 & 117 & 10 & 8.5 \\
\midrule
\multicolumn{5}{l}{\textit{GT says nothing, contradicted}} \\
\hspace{1em}Required & 15 & 15 & 2 & 13.3 \\
\hspace{1em}Constraint & 124 & 124 & 3 & 2.4 \\
\hspace{1em}Status & 153 & 153 & 23 & 15.0 \\
\hspace{1em}Field & 10 & 10 & 1 & 10.0 \\
\hspace{1em}Authentication & 44 & 44 & 0 & 0.0 \\
\midrule
\multicolumn{5}{l}{\textit{GT says nothing, confirmed}} \\
\hspace{1em}Registry route & 161 & 161 & 5 & 3.1 \\
\hspace{1em}Invented route & 214 & 214 & 76 & 35.5 \\
\hspace{1em}Parameter & 268 & 268 & 49 & 18.3 \\
\hspace{1em}Required & 25 & 25 & 1 & 4.0 \\
\hspace{1em}Constraint & 355 & 355 & 129 & 36.3 \\
\hspace{1em}Authentication & 206 & 206 & 0 & 0.0 \\
\hspace{1em}Status$^*$ & 1084 & 46 & 4 & 8.7 \\
\hspace{1em}Field$^*$ & 579 & 27 & 0 & 0.0 \\
\midrule
All & 10254 & 1865 & 306 & 16.4 \\
\bottomrule
\end{tabular}
\end{table}
The first part of the table compares the Scorer with the GT. When the Scorer and the GT disagree, the human evaluation shows the GT to be wrong in 91.5\% (107/117) of the disagreements. The Scorer and the GT agree on 6899 outcomes. The human evaluated 100 of these outcomes and found the Scorer wrong on only 3\% (3/100). This shows that when the Scorer and the GT agree, both are almost always right.

Table~\ref{tab:human-evaluation} also shows the other two parts of this evaluation, which are the outcomes that the GT says nothing about. When the GT says nothing and the Scorer contradicts a claim, it is wrong on 8.4\% (29/346) of the human evaluated outcomes. When the GT says nothing and the Scorer confirms a claim, it is wrong on 20.3\% (264/1302). So, most errors are in claims that the GT cannot settle. Over the whole human evaluation, the Scorer is wrong on 16.4\% (306/1865) of the evaluated outcomes.

The 16.4\% overstates the error of the Scorer. The human evaluated every outcome where errors were likely and sampled the rows marked with an asterisk. If we count the sampled outcomes for the outcomes they stand for and the non-sampled outcomes as they are, the Scorer is wrong on about 6.4\% (660/10,254) of all outcomes. The samples are small, so the rate could lie between 3.7\% and 11.5\%, if we take each sample at the ends of its 95\% interval. A second human re-evaluated 5\% of the evaluated outcomes, drawn at random, and agreed on 91 of the 95 outcomes, with a Cohen's kappa of 0.89.

Most errors come from three limits of deciding a claim with requests. First, the Scorer only sees the cases its requests exercise. For example, if a service answers 401 when a token has expired and no request sends an expired token, the Scorer would never see the 401 and contradict it. Second, the Scorer cannot see why the service accepted or refused a request. It confirms a parameter when the service accepts a request that carries it, even if the service ignores the parameter, and it confirms a constraint when the service refuses a request that breaks it, even if the service refused for a different reason. Third, the Scorer confirms an invented route when a redirect, a catch-all or another route answers in place of a 404 for that route. Together, these account for 90.5\% (277/306) of the outcomes where the human evaluation finds the Scorer wrong.
\subsubsection*{\textbf{RQ0 Answer}}
The Scorer is wrong only on 6.4\% (660/10,254) of its outcomes. When it disagrees with the GT, it is right in 91.5\% (107/117) of the cases. Most of the outcomes where the human evaluation finds the Scorer wrong come from cases its requests do not exercise, requests the service accepts or refuses for another reason, and invented routes that the service answers somehow in place of a 404. Since the Scorer is more correct than the ground truth and as it allows evaluation without manual creation of ground truth, we use it later to measure every specification against the running service.

\subsection{\rqOasisvsServices}
\label{sec:results:rq1}
Table~\ref{tab:services} shows the results of running OASIS for 30 services and the evaluation metrics presented in Section \ref{sec:metrics}.
We see that 19 services have 1.000 route recall meaning that OASIS captured all the routes present in the route registry, and the mean route recall over all services is 0.989. 

OASIS has a precision of 0.967 at a tested share of 0.585, so the running service confirms nearly all of the claims that the Scorer could send a request for. Precision ranges from 0.908 for Laravel to 0.988 for Spring among the frameworks with five services. Precision is at least 0.941 on every service other than Koel. Koel is the only outlier due to the way the service replies. Almost all of its contradicted claims come from its Subsonic API under \texttt{/rest}. This API reports failed requests inside the body of a 200 status code reply. The Scorer reads these failures as success because of the status code, thus the refusals act as an acceptance.

Table~\ref{tab:static} shows how the results evolve from the Static Draft of stage 1 to the final OpenAPI Specification of stage 3. We see that the Static Draft already reaches nearly the same mean route recall of 0.990 as the final version. Therefore, the routes mainly come from stage 1. The largest change from the Static Draft to the final version is in the responses, meaning their status codes and fields. The field recall rises from 0.738 to 0.980 and the status recall from 0.913 to 0.947. Field recall rises on 24 services and status recall on 20. These gains are the largest where the source does not declare any type for the response body, so the Static Draft sees nothing for these. The largest gains are on Koel, where it goes from 0.014 to 1.000, NodeBB, where it goes from 0.005 to 0.962 and Akaunting, where it goes from 0.123 to 0.992. Precision, tested share and GT agreement also rise but only a little.

We note that OASIS invents very few routes, 96.7 on average over the three runs. The largest share of these is on NodeBB, 42.7 per run, and the human evaluation of NodeBB's invented routes in RQ0 finds 22 out of 26 to be real routes which the route registry leaves out.

Between the three runs, the median of the range over the 30 services is 0.006 in precision and 0.004 in status recall. Route recall does not differ on 25 of the 30 services and field recall does not differ on 20. The widest range is in field recall, 0.144 on Gramps-Web-API, due to different requests and replies drawn by the agent. Over the three runs, the mean cost of a run over the 30 services is 21.61~USD, which comes to about one cent per route for all three stages. Although the timing of the runs can vary due to model and provider latency, we note that the runs take a median of 4.4 minutes per service and at most 62 minutes on NodeBB. The limits that are set in stage 2 are rarely exhausted. About 2\% of the Agent Sessions used all 25 HTTP calls, and none reached the 60 model calls or 900 seconds limit.

\subsubsection*{\textbf{RQ1 Answer}}
OASIS has a route recall of 0.989 and a precision of 0.967 at a tested share of 0.585. Precision is at least 0.941 on every service but Koel. Its API cost is about one cent per route. Stage 1 finds the routes, and probing adds what the source does not show. Probing raises field recall from 0.738 to 0.980 and status recall from 0.913 to 0.947.

\begin{table}[t]
\centering
\caption{OASIS Static Draft (after Stage 1) and final (after Stage 3) OpenAPI Specification, means over the 30 services and 3 runs, with tested share over all claims and total invented routes.}
\label{tab:static}
\footnotesize
\setlength{\tabcolsep}{2pt}
\begin{tabular}{l ccc cc r c}
\toprule
 & Precision & \shortstack{Tested\\share} & \shortstack{Route\\recall} & \shortstack{Status\\recall} & \shortstack{Field\\recall} & \shortstack{Invented\\routes} & \shortstack{GT\\agreement} \\
\midrule
Static Draft & 0.955 & 0.545 & 0.990 & 0.913 & 0.738 & 110.3 & 0.869 \\
Final & \textbf{0.967} & 0.585 & 0.989 & \textbf{0.947} & \textbf{0.980} & \textbf{96.7} & \textbf{0.877} \\
\bottomrule
\end{tabular}
\end{table}

\begin{table*}[t]
\centering
\caption{The 30 services per framework. Means of 3 OASIS runs on each service with GPT-5.6 Luna. The last row gives means over services, totals for operations, invented routes and cost, and tested share over all claims.}
\label{tab:services}
\setlength{\tabcolsep}{3pt}
\newcommand{\fw}[1]{{\scriptsize #1}}
\newcommand{\fwc}[1]{\raisebox{-.5\normalbaselineskip}[0pt][0pt]{\scriptsize #1}}
\begin{tabular}{l l rr c r cc c cc r r}
\toprule
 & Service & LOC & Routes & GT & Operations & Precision & \shortstack{Tested\\share} & \shortstack{Route\\recall} & \shortstack{Status\\recall} & \shortstack{Field\\recall} & \shortstack[r]{Invented\\routes} & \shortstack[r]{Cost\\(USD)} \\
\midrule
 & BlogAPI & 3\,718 & 54 & -- & 53 & 0.994 & 0.595 & 0.988 & 0.928 & 0.992 & 0.0 & 0.61 \\
\fwc{Spring} & HTTP-Patch-Spring & 1\,008 & 6 & -- & 6 & 0.996 & 0.929 & 1.000 & 0.945 & 1.000 & 0.0 & 0.03 \\
\fwc{(Java)} & Market & 10\,372 & 13 & \checkmark & 13 & 0.981 & 0.754 & 1.000 & 1.000 & 1.000 & 0.0 & 0.13 \\
 & OOPS-Proxy-Print & 5\,360 & 71 & \checkmark & 72 & 0.972 & 0.478 & 1.000 & 0.894 & 0.982 & 1.0 & 0.65 \\
 & PetClinic & 5\,382 & 40 & \checkmark & 42 & 0.998 & 0.793 & 1.000 & 1.000 & 1.000 & 2.0 & 0.49 \\
\midrule
 & Digdag & 75\,721 & 40 & \checkmark & 40 & 0.967 & 0.497 & 0.992 & 0.953 & 0.963 & 0.3 & 0.54 \\
\fwc{Jersey} & Features-Service & 1\,313 & 18 & \checkmark & 18 & 1.000 & 0.381 & 1.000 & 1.000 & 1.000 & 0.0 & 0.10 \\
\fwc{(Java)} & RESTCountries & 1\,369 & 27 & -- & 27 & 1.000 & 0.556 & 1.000 & 1.000 & 1.000 & 0.3 & 0.22 \\
 & Scout-API & 6\,573 & 49 & \checkmark & 49 & 0.981 & 0.360 & 1.000 & 0.799 & 0.960 & 0.0 & 0.49 \\
 & Swagger-Petstore & 1\,083 & 19 & \checkmark & 20 & 0.963 & 0.686 & 1.000 & 0.929 & 1.000 & 0.7 & 0.08 \\
\midrule
 & Conduit-RealWorld & 3\,427 & 19 & -- & 19 & 0.974 & 0.593 & 1.000 & 1.000 & 1.000 & 0.0 & 0.13 \\
\fwc{Express} & NodeBB & 101\,982 & 523 & \checkmark & 496 & 0.972 & 0.687 & 0.866 & 0.813 & 0.962 & 42.7 & 3.56 \\
\fwc{(JS/TS)} & OOPS-Accountill & 5\,488 & 24 & \checkmark & 24 & 0.951 & 0.530 & 1.000 & 0.826 & 0.986 & 0.0 & 0.13 \\
 & ProShop & 3\,114 & 23 & \checkmark & 25 & 0.971 & 0.562 & 1.000 & 1.000 & 1.000 & 2.0 & 0.15 \\
 & Oodikone & 51\,496 & 99 & -- & 104 & 0.978 & 0.638 & 1.000 & 1.000 & 0.989 & 4.7 & 0.72 \\
\midrule
 & AISysRev & 9\,190 & 29 & \checkmark & 29 & 1.000 & 0.583 & 1.000 & 1.000 & 1.000 & 0.0 & 0.12 \\
\fw{Flask/} & CTFd & 55\,230 & 215 & \checkmark & 215 & 0.982 & 0.694 & 0.974 & 0.861 & 1.000 & 5.7 & 2.35 \\
\fw{FastAPI} & Gramps-Web-API & 33\,029 & 168 & \checkmark & 183 & 0.978 & 0.522 & 0.970 & 0.802 & 0.832 & 19.7 & 2.47 \\
\fw{(Python)} & httpbin & 2\,001 & 82 & \checkmark & 81 & 0.960 & 0.640 & 0.988 & 0.944 & 0.988 & 0.0 & 0.76 \\
 & Lin-CMS & 4\,315 & 31 & \checkmark & 34 & 0.989 & 0.627 & 1.000 & 0.994 & 1.000 & 2.7 & 0.33 \\
\midrule
 & Akaunting & 171\,419 & 121 & -- & 127 & 0.995 & 0.508 & 0.992 & 0.993 & 0.992 & 7.3 & 1.34 \\
\fwc{Laravel} & F1amy-RealWorld & 3\,209 & 21 & -- & 21 & 0.962 & 0.384 & 1.000 & 1.000 & 1.000 & 0.0 & 0.18 \\
\fwc{(PHP)} & Gothinkster-RealWorld & 2\,500 & 21 & -- & 21 & 0.941 & 0.617 & 1.000 & 0.960 & 1.000 & 0.0 & 0.15 \\
 & Koel & 80\,455 & 299 & -- & 301 & 0.698 & 0.506 & 0.993 & 0.983 & 1.000 & 4.3 & 3.36 \\
 & OOPS-Laravel-Blog & 6\,097 & 23 & \checkmark & 23 & 0.943 & 0.429 & 1.000 & 1.000 & 1.000 & 0.0 & 0.19 \\
\midrule
 & Gin-Admin & 5\,449 & 27 & \checkmark & 27 & 0.991 & 0.558 & 0.963 & 0.907 & 1.000 & 1.0 & 0.37 \\
\fwc{Gin} & Gotify-Server & 13\,057 & 41 & \checkmark & 44 & 0.992 & 0.591 & 0.976 & 0.995 & 0.744 & 2.0 & 0.62 \\
\fwc{(Go)} & OOPS-RealWorld & 1\,173 & 19 & \checkmark & 19 & 0.979 & 0.523 & 1.000 & 1.000 & 1.000 & 0.0 & 0.19 \\
 & PaoPao-CE & 28\,109 & 71 & -- & 71 & 0.954 & 0.846 & 1.000 & 1.000 & 1.000 & 0.3 & 0.97 \\
 & RealWorld-Go & 2\,469 & 20 & \checkmark & 20 & 0.951 & 0.552 & 0.983 & 0.874 & 1.000 & 0.0 & 0.19 \\
\midrule
\multicolumn{2}{l}{All} & 695\,108 & 2213 & 20 & 2223 & 0.967 & 0.585 & 0.989 & 0.947 & 0.980 & 96.7 & 21.61 \\
\bottomrule
\end{tabular}
\end{table*}

\subsection{\rqOasisvsBaselines}
\label{sec:results:rq2}

Table~\ref{tab:baselines} compares each baseline alongside OASIS. Each baseline was run against a subset of services because a baseline runs only on the supported frameworks. We limit LRASGen to the two largest services of each framework it supports due to its cost. OASIS produces a specification for all 30 services. Respector runs on 10, Speculate on 10, ExpressO on 5, LRASGen on 9 and OOPS on 30. Ten of these 64 runs produced no specification, due to a crash, a timeout or a missing entry file.

Table~\ref{tab:baselines} shows that OASIS leads all five baselines on claim F1. OOPS is the closest baseline to OASIS with a claim F1 of 0.827 against 0.957. Over the entire comparison, we see that OASIS has a higher claim F1 than 52 of the 54 baseline specifications. Every baseline drops below a claim F1 of 0.14 on at least one service, where it misses most of the registered routes. OASIS stays at or above 0.834 on every service.

\begin{table*}[t]
\centering
\caption{Each baseline beside OASIS on the services where the baseline produced a specification. Values are means over these services, with tested share over all claims and totals for invented routes and cost.}
\label{tab:baselines}
\setlength{\tabcolsep}{4pt}
\begin{tabular}{l r cccccc r c r}
\toprule
Specification & Services & \shortstack{Claim\\F1} & Precision & \shortstack{Tested\\share} & \shortstack{Route\\recall} & \shortstack{Status\\recall} & \shortstack{Field\\recall} & \shortstack{Invented\\routes} & \shortstack{GT\\agreement} & \shortstack{Cost\\(USD)} \\
\midrule
Respector & 8 & 0.536 & 0.942 & 0.479 & 0.875 & 0.432 & 0.221 & 5 & 0.500 & 0.00 \\
OASIS & 8 & \textbf{0.971} & \textbf{0.982} & 0.495 & \textbf{0.999} & \textbf{0.916} & \textbf{0.986} & \textbf{2.3} & \textbf{0.885} & 2.24 \\
\midrule
Speculate & 10 & 0.699 & 0.971 & 0.416 & 0.807 & 0.530 & 0.643 & 5 & 0.512 & 2.29 \\
OASIS & 10 & \textbf{0.971} & \textbf{0.985} & 0.553 & \textbf{0.998} & \textbf{0.924} & \textbf{0.987} & \textbf{4.3} & \textbf{0.867} & 3.34 \\
\midrule
ExpressO & 3 & 0.277 & \textbf{0.978} & 0.500 & 0.725 & 0.434 & 0.000 & 12 & 0.373 & 0.00 \\
OASIS & 3 & \textbf{0.955} & 0.965 & 0.553 & \textbf{1.000} & \textbf{0.940} & \textbf{0.999} & \textbf{2.0} & \textbf{0.938} & 0.41 \\
\midrule
LRASGen & 8 & 0.508 & 0.824 & 0.383 & 0.684 & 0.502 & 0.000 & 405 & 0.447 & 153.20 \\
OASIS & 8 & \textbf{0.931} & \textbf{0.981} & 0.593 & \textbf{0.974} & \textbf{0.868} & \textbf{0.955} & \textbf{69.3} & \textbf{0.800} & 10.79 \\
\midrule
OOPS & 25 & 0.827 & 0.969 & 0.455 & 0.882 & 0.746 & 0.837 & 217 & 0.877 & 22.95 \\
OASIS & 25 & \textbf{0.957} & \textbf{0.978} & 0.587 & \textbf{0.993} & \textbf{0.932} & \textbf{0.976} & \textbf{41.7} & 0.877 & 13.09 \\
\bottomrule
\end{tabular}
\end{table*}

\begin{table}[t]
\centering
\caption{F1 per kind of claim on the services of Table~\ref{tab:baselines}.}
\label{tab:kinds}
\footnotesize
\setlength{\tabcolsep}{2.5pt}
\begin{tabular}{l ccccccc}
\toprule
Specification & Route & Param. & Req. & Constr. & Auth. & Status & Field \\
\midrule
Respector & 0.874 & 0.825 & 0.000 & 0.513 & 0.086 & 0.526 & 0.270 \\
OASIS & \textbf{0.994} & \textbf{0.987} & \textbf{0.917} & \textbf{0.938} & \textbf{0.948} & \textbf{0.963} & \textbf{0.996} \\
\midrule
Speculate & 0.822 & 0.795 & 0.611 & 0.738 & 0.276 & 0.645 & 0.673 \\
OASIS & \textbf{0.993} & \textbf{0.973} & \textbf{0.942} & \textbf{0.936} & \textbf{0.957} & \textbf{0.966} & \textbf{0.996} \\
\midrule
ExpressO & 0.770 & 0.354 & 0.000 & 0.000 & 0.000 & 0.559 & 0.000 \\
OASIS & \textbf{1.000} & \textbf{1.000} & \textbf{0.881} & \textbf{0.820} & \textbf{0.959} & \textbf{0.996} & \textbf{0.999} \\
\midrule
LRASGen & 0.757 & 0.706 & 0.359 & 0.523 & 0.468 & 0.638 & 0.000 \\
OASIS & \textbf{0.950} & \textbf{0.910} & \textbf{0.959} & \textbf{0.794} & \textbf{0.961} & \textbf{0.945} & \textbf{0.993} \\
\midrule
OOPS & 0.890 & 0.908 & 0.759 & 0.723 & 0.680 & 0.839 & 0.883 \\
OASIS & \textbf{0.984} & \textbf{0.973} & \textbf{0.967} & \textbf{0.847} & \textbf{0.952} & \textbf{0.966} & \textbf{0.994} \\
\bottomrule
\end{tabular}
\end{table}

A specification that has fewer claims can be precise on those few. Across all five baselines, OASIS has a higher precision than four baselines. The only exception is ExpressO. However, we see that requests only decide 160 of its claims, against 832 of OASIS, and its claim F1 is 0.277 against 0.955. OASIS has the most tested share amongst all baselines, and over the 54 comparisons, requests decide 57.6\% of its claims. OASIS has a higher route recall than every baseline and fewer invented routes. When the Scorer creates requests for these invented routes, the service answers 74\% of the invented routes of OASIS against 31\% for the baselines altogether.

OASIS is the only tool that writes descriptions and request examples in the specifications, and it writes a summary and a description for 0.89 to 1.00 of all its operations. These are in natural language and are written using information the LLM saw during route documentation. Speculate is the only other tool which writes summaries for operations.

OASIS costs 43\% less than OOPS and 93\% less than LRASGen. LRASGen spends 121.14 of its 153.20~USD on NodeBB, and without NodeBB OASIS still costs 77\% less.

GT agreement for OASIS is also high. Comparing OASIS to OOPS on this agreement reveals that both tools agree nearly equally with the ground truths. Seventeen of the 25 services of OOPS have a ground truth in the measured comparison here, and on six of them OASIS loses to OOPS. On five of these six, OASIS documents more than the ground truth and the agreement counts the extra against it. For example, the biggest loss in agreement is on CTFd, where the ground truth covers only \texttt{/api/v1} and leaves out 79 of the 215 registered routes, and OASIS documents 74 or 75 of these in each run. 

Table~\ref{tab:kinds} splits claim F1 into its seven kinds of claims. We see that OASIS has the higher F1 on every kind against every baseline. The biggest gap is on authentication and fields. OASIS has an F1 above 0.94 on both of these kinds, while the nearest baseline has 0.680 for authentication and 0.883 for fields. This is because a tool that does static source analysis has to infer these claims. Authentication is configured in the middleware, and response bodies are assembled at runtime. Dynamic validation of the specification with probing observes both. OASIS also has a higher status recall and a higher field recall than every baseline. The baselines also list fewer response fields since the source rarely names them. Only 20 of the 229 operations of Respector and only 16 of the 1015 of LRASGen name any field in a success reply.

\subsubsection*{\textbf{RQ2 Answer}}
OASIS has a higher claim F1 than all five baselines and than 52 of the 54 baseline specifications. 
The gap is widest on authentication and fields, which a reader of the source has to infer. OASIS also has a higher route recall and fewer invented routes than the baselines. OASIS costs less than OOPS and LRASGen, and it is the only tool with a specification for every service.

\subsection{\rqOasisvsModels}
\label{sec:results:rq3}

\begin{table}[t]
\centering
\caption{OASIS with seven models on the seven RQ3 services.}
\label{tab:model-set}
\footnotesize
\setlength{\tabcolsep}{2pt}
\begin{tabular}{l cc c cc r r}
\toprule
Model & \shortstack{Preci-\\sion} & \shortstack{Tested\\share} & \shortstack{Route\\recall} & \shortstack{Status\\recall} & \shortstack{Field\\recall} & \shortstack{Invented\\routes} & \shortstack{Cost\\(USD)} \\
\midrule
GPT-5.6 Luna & \textbf{0.939} & \textbf{0.622} & \textbf{0.978} & 0.868 & \textbf{0.957} & \textbf{47} & 11.90 \\
Gemini 3.1 Flash Lite & 0.915 & 0.508 & 0.965 & 0.781 & 0.863 & 263 & 8.61 \\
DeepSeek V4 Flash & 0.907 & 0.513 & 0.968 & 0.853 & 0.884 & 217 & 10.47 \\
gpt-oss-120b & 0.917 & 0.548 & 0.972 & 0.816 & 0.889 & 64 & 7.30 \\
GLM 5.3 Flash & 0.931 & 0.544 & 0.970 & \textbf{0.913} & 0.884 & 48 & 34.42 \\
Qwen3 235B & 0.865 & 0.439 & 0.957 & 0.763 & 0.871 & 128 & 10.33 \\
Llama 4 Scout & 0.802 & 0.353 & 0.916 & 0.704 & 0.622 & 637 & \textbf{5.12} \\
\bottomrule
\end{tabular}
\end{table}

We measured all seven models on the largest service of each framework by route count (OOPS-Proxy-Print, Scout-API, NodeBB, CTFd, AISysRev, Koel and PaoPao-CE)
and Table~\ref{tab:model-set} gives the results. OASIS performance in terms of finding routes (recall) has small variation between the models as the range of the models is 0.916 to 0.978.

In terms of precision and tested share, the models vary more. Table~\ref{tab:model-set} shows that precision ranges from 0.802 to 0.939 and tested share from 0.353 to 0.622.

Status recall ranges from 0.704 to 0.913, and field recall from 0.622 to 0.957, while the invented routes range from 47 to 637. GLM 5.3 Flash has the highest value of status recall and GPT-5.6 Luna is the second. GPT-5.6 Luna leads on precision, route recall, tested share and field recall. Llama 4 Scout is the weakest on every measure except cost.

Llama 4 Scout is the cheapest model at 5.12~USD. GLM 5.3 Flash costs the most at 34.42~USD, because 98\% of its output is reasoning. gpt-oss-120b comes close to GPT-5.6 Luna on route recall, at 61\% of the cost of GPT-5.6 Luna. Even with Llama 4 Scout, OASIS still has a route recall of 0.916 and a precision of 0.802. This shows that the pipeline holds up with a weak model.

\subsubsection*{\textbf{RQ3 Answer}}
Route recall stays between 0.916 and 0.978 and precision between 0.802 and 0.939 for the seven models. The model matters more for precision and tested share than for the routes. GPT-5.6 Luna leads on most measures.

\section{Threats to Validity}
\label{sec:threats}

A threat to construct validity is that the Scorer is our own and due to the way it constructs the request, it shares some rules with stage 3 of OASIS. Both mark a field required when leaving it out draws a 4xx, and both take status codes and fields from replies. This is hard to avoid, since a running service shows a required field only by refusing a request that does not carry it, and shows the status codes and fields only in its replies.
With the human outcomes of RQ0 in place of the Scorer, OASIS still leads 51 of the 54 baseline specifications.
Another threat to construct validity is that OASIS documents the service as it runs, so a service's defect becomes a specification's claim. Black-box testing tools that use this specification will report deviations from the documented behavior and miss the defects it documents, but they still report server errors as failures~\cite{arcuri2021evomaster,atlidakis2019restler,hatfielddodds2022schemathesis}. This issue also exists for a specification read from the source, since the source holds the same defects.

A threat to internal validity is the variation of LLM output, even at temperature zero. RQ1 shows that OASIS results vary little between the three runs, with a median range of 0.006 in precision, 0.004 in status recall, and 0 in route and field recall. Each baseline runs once. A difference of a few hundredths between two models therefore might not be a reliable ranking.

A threat to external validity is that popular services and their specifications may be in the models' training data. This should not affect the comparison against prior works in Section \ref{sec:results:rq2} as we run them on the same model as OASIS. However, prior works might perform better with other models, yet, comparing all prior works across all models would increase research costs.

\section{Conclusion}
\label{sec:conclusion}

REST APIs widely use OpenAPI specifications. Documentation, client code generation and API testing all trust the claims in the specification. We present OASIS, an OpenAPI specification generation tool that drafts a specification from the source code and then checks and repairs it against the running service. We evaluate OASIS on 30 services in seven frameworks and five languages, and we score every specification against the running service with our own Scorer instead of a ground truth. The Scorer is shown to be superior against the given ground truth in the human evaluation. The Scorer is correct on about 94\% of its outcomes. OASIS has a route recall of 0.989 and a precision of 0.967 at a tested share of 0.585, for about one cent of API cost per registered route. OASIS beats its baselines in claim F1 in 96\% of the specification tasks.
OASIS also does well with cheap models. OASIS can help reduce the manual effort in specification generation and validate specifications against the running service to show what the source alone cannot. For future work, we want to explore an agent for specification evaluation, and probing with other roles and more data to raise the tested share. We release OASIS, the Scorer, the benchmark, the route registries and the data of the human evaluation at \url{https://figshare.com/s/112f2f999856d11691f2}.

\section*{Acknowledgments}
The authors used Claude Code (Anthropic) to assist with some parts of the OASIS implementation and analysis scripts, along with minor language edits in the paper. The authors verified every output and take full responsibility for the content. 

\bibliographystyle{IEEEtran}
\bibliography{oasis}

\end{document}